\documentclass{elsart}
\usepackage{graphicx}
\usepackage{times}
\usepackage{amssymb}

\begin{document}

\begin{frontmatter}


\title{The STAR Time Projection Chamber:
A Unique Tool for Studying High Multiplicity Events at RHIC}

\author[UCD]{M.~Anderson}, 
\author[LBL]{J.~Berkovitz}, 
\author[BNL]{W.~Betts},
\author[LBL]{R.~Bossingham}, 
\author[LBL]{F.~Bieser}, 
\author[BNL]{R.~Brown},
\author[LBL]{M.~Burks},
\author[BNL]{M.~Calder\'{o}n~de~la~Barca~S\'{a}nchez},
\author[UCD]{D.~Cebra}, 
\author[Creighton]{M.~Cherney}, 
\author[Creighton]{J.~Chrin}, 
\author[LBL]{W.R.~Edwards}, 
\author[UCLA]{V.~Ghazikhanian},
\author[LBL]{D.~Greiner}, 
\author[Purdue]{M.~Gilkes}, 
\author[LBL]{D.~Hardtke},
\author[UW]{G.~Harper},
\author[LBL]{E.~Hjort},
\author[UCLA]{H.~Huang}, 
\author[UCLA]{G.~Igo}, 
\author[LBL]{S.~Jacobson}, 
\author[Kent]{D.~Keane}, 
\author[LBL]{S.R.~Klein}, 
\author[LBL]{G.~Koehler},
\author[PNPI]{L.~Kotchenda}, 
\author[Yale]{B.~Lasiuk}, 
\author[BNL]{A.~Lebedev}, 
\author[Creighton]{J.~Lin},
\author[Ohio]{M.~Lisa}, 
\author[LBL]{H.S.~Matis}, 
\author[LBL]{J.~Nystrand},
\author[BNL]{S.~Panitkin}, 
\author[Creighton]{D.~Reichold}, 
\author[LBL]{F.~Retiere}, 
\author[LBL]{I.~Sakrejda}, 
\author[LBL]{K.~Schweda}, 
\author[LBL]{D.~Shuman},
\author[LBL]{R.~Snellings},
\author[BNL,LBL]{N.~Stone}, 
\author[Purdue]{B.~Stringfellow}, 
\author[LBL]{J.H.~Thomas}, 
\author[UW]{T.~Trainor}, 
\author[UCLA]{S.~Trentalange}, 
\author[LBL]{R.~Wells},
\author[UCLA]{C.~Whitten}, 
\author[LBL]{H.~Wieman}, 
\author[LBL,UCLA]{E.~Yamamoto}, 
\author[Kent]{W.~Zhang}

\address[BNL] {Brookhaven National Laboratory, Upton, NY 11973,  USA}
\address[Creighton] {Creighton University, Omaha, NE 68178,  USA}
\address[Kent] {Kent State University, Kent, Ohio 44242,  USA}
\address[LBL] {Lawrence Berkeley National Laboratory, Berkeley, CA  94720,  USA}
\address[Ohio] {Ohio State University, Columbus, OH  43210,  USA}
\address[Purdue] {Purdue University, West Lafayette, IN  47907,  USA}
\address[PNPI] {St-Petersburg Nuclear Physics Institute, Gatchina 188350, Russia}
\address[UCD] {University of California at Davis, Davis, CA 95616,  USA}
\address[UCLA] {University of California at Los Angeles, Los Angeles, CA  90095, USA}
\address[UW] {University of Washington, Seattle, WA  98195,  USA}
\address[Yale] {Yale University, New Haven, CT 06520,  USA}

\begin{abstract}
  
  The STAR Time Projection Chamber  (TPC) is used to record collisions
  at  the Relativistic  Heavy  Ion  Collider (RHIC).  The  TPC is  the
  central  element  in  a   suite  of  detectors  that  surrounds  the
  interaction vertex.   The TPC provides complete  coverage around the
  beam-line,  and  provides complete  tracking  for charged  particles
  within  $\pm$ 1.8  units  of pseudo-rapidity  of the  center-of-mass
  frame.  Charged  particles with momenta  greater than 100  MeV/c are
  recorded.  Multiplicities  in excess of  3,000 tracks per  event are
  routinely reconstructed  in the  software. The TPC  measures 4  m in
  diameter by 4.2 m long, making it the largest TPC in the world.

\end{abstract}

\begin{keyword}
Detectors \sep TPC \sep Time Projection Chambers \sep Drift Chamber \sep Heavy Ions
\PACS  29.40.-n \sep 29.40.Gx
\end{keyword}

\end{frontmatter}


\section{Introduction}

The Relativistic  Heavy Ion Collider  (RHIC) is located  at Brookhaven
National Laboratory.  It accelerates heavy  ions up to a top energy of
100 GeV per nucleon, per beam.   The maximum center of mass energy for
Au+Au  collisions  is $\sqrt{s_{NN}}  =  200$  GeV  per nucleon.  Each
collision produces a large number of charged particles. For example, a
central Au-Au collision will  produce more than 1000 primary particles
per  unit of  pseudo-rapidity.   The average  transverse momentum  per
particle is about 500 MeV/c.  Each collision also produces a high flux
of secondary particles that are  due to the interaction of the primary
particles with  the material in the  detector, and the  decay of short
lived  primaries.   These  secondary  particles must  be  tracked  and
identified along with the primary particles in order to accomplish the
physics  goals of  the experiment.   Thus,  RHIC is  a very  demanding
environment in which to operate a detector.

The  STAR  detector\cite{CDR,CDRU,web} uses  the  TPC  as its  primary
tracking  device\cite{TPC1,TPC2}.   The  TPC  records  the  tracks  of
particles,  measures their  momenta, and  identifies the  particles by
measuring  their  ionization  energy  loss ($dE/dx$).  Its  acceptance
covers $\pm  1.8$ units of pseudo-rapidity through  the full azimuthal
angle  and  over the  full  range  of  multiplicities.  Particles  are
identified  over a momentum  range from  100 MeV/c  to greater  than 1
GeV/c, and momenta are measured over a range of 100 MeV/c to 30 GeV/c.

\begin{figure}[htb]
\includegraphics[width=14cm]{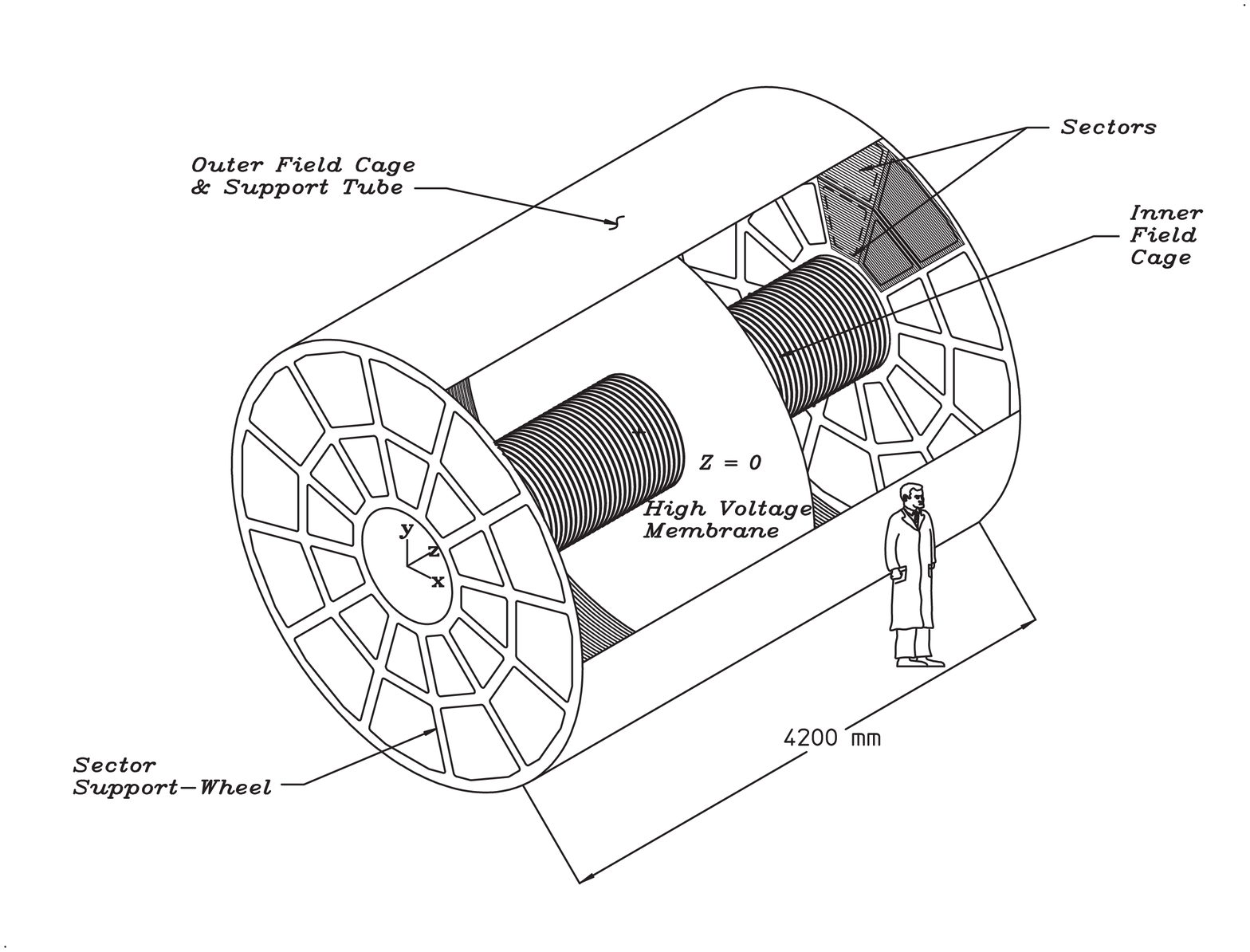}
\caption{The STAR TPC surrounds a beam-beam interaction region at RHIC. The
  collisions take place near the center of the TPC.}
\label{tpcman}
\end{figure}

The STAR TPC is shown schematically in Fig. \ref{tpcman}. It sits in a
large solenoidal magnet that  operates at 0.5 T\cite{magnet}.  The TPC
is 4.2 m long  and 4 m in diameter. It is an empty  volume of gas in a
well  defined, uniform,  electric field  of $\approx$  135  V/cm.  The
paths of primary ionizing particles passing through the gas volume are
reconstructed  with   high  precision  from   the  released  secondary
electrons  which drift  to the  readout end  caps at  the ends  of the
chamber.  The  uniform electric field  which is required to  drift the
electrons is defined by a thin conductive Central Membrane (CM) at the
center of the TPC, concentric field-cage cylinders and the readout end
caps.    Electric   field   uniformity   is   critical   since   track
reconstruction precision  is  sub-millimeter and electron  drift paths
are up to 2.1 meters.

The readout system is based on Multi-Wire Proportional Chambers (MWPC)
with  readout pads.   The  drifting electrons  avalanche  in the  high
fields at the 20 $\mu$m anode wires providing an amplification of 1000
to  3000.   The  positive  ions  created in  the  avalanche  induce  a
temporary image charge  on the pads which disappears  as the ions move
away  from  the  anode  wire.   The  image charge  is  measured  by  a
preamplifier/shaper/waveform  digitizer  system.   The induced  charge
from  an  avalanche is  shared  over  several  adjacent pads,  so  the
original track position can be  reconstructed to a small fraction of a
pad width.  There are a total of 136,608 pads in the readout system.

The TPC is filled with P10 gas (10\% methane, 90\% argon) regulated at
2 mbar  above atmospheric pressure\cite{gas}.  This gas  has long been
used in TPCs.   It's primary attribute is a  fast drift velocity which
peaks at a low electric field.   Operating on the peak of the velocity
curve  makes  the  drift  velocity  stable and  insensitive  to  small
variations   in  temperature  and   pressure.   Low   voltage  greatly
simplifies the field cage design.

The design and specification strategy  for the TPC have been guided by
the limits of the gas and  the financial limits on size.  Diffusion of
the drifting  electrons and their limited number  defines the position
resolution.  Ionization fluctuations and finite track length limit the
$dE/dx$  particle  identification.   The  design  specifications  were
adjusted accordingly  to limit  cost and complexity  without seriously
compromising  the  potential   for  tracking  precision  and  particle
identification.

Table \ref{majorparam}  lists some basic parameters for  the STAR TPC.
The measured TPC performance  has generally agreed with standard codes
such as MAGBOLTZ\cite{Magboltz} and GARFIELD\cite{Garfield}.  Only for
the  most  detailed studies  has  it  been  necessary to  make  custom
measurements of  the electrostatic or  gas parameters (e.g.  the drift
velocity in the gas).

\begin{table}[htb]
\begin{tabular}{|l|l|l|}
\hline
Item &  Dimension  &    Comment \\
\hline
Length of the TPC  &    420 cm   & Two halves, 210 cm long \\
Outer Diameter of the Drift Volume &    400 cm & 200 cm radius \\
Inner Diameter of the Drift Volume &    100 cm & 50 cm radius \\
Distance: Cathode to Ground Plane  &    209.3 cm &      Each side \\
Cathode  & 400 cm diameter &    At the center of the TPC \\
Cathode Potential &     28 kV & Typical \\
Drift Gas &     P10     & 10\% methane, 90\% argon \\ 
Pressure &      Atmospheric + 2 mbar &  Regulated at 2 mbar above Atm. \\
Drift Velocity & 5.45 cm / $\mu$s &     Typical \\
Transverse Diffusion ($\sigma$)   & $230 \mu m / \sqrt{cm}$ 
& 140 V/cm \& 0.5 T \\
Longitudinal Diffusion ($\sigma$) & $360 \mu m / \sqrt{cm}$ 
& 140 V/cm \\
Number of Anode Sectors  &      24 &    12 per end \\
Number of Pads		&	136,608 &      \\
Signal to Noise Ratio & 20 : 1  & \\  
Electronics Shaping Time &      180 ns &        FWHM \\
Signal Dynamic Range &  10 bits & \\    
Sampling Rate & 9.4 MHz & \\     
Sampling Depth &        512 time buckets &      380 time buckets typical \\
Magnetic Field &         0,  $\pm 0.25$ T, $\pm 0.5$  T & Solenoidal \\
\hline
\end{tabular}
\vspace{0.2cm}
\caption{Basic parameters for the STAR TPC and its associated hardware.}
\label{majorparam}
\end{table}

\section{Cathode and Field Cage}

The uniform electric  field in the TPC is  defined by establishing the
correct  boundary conditions with  the parallel  disks of  the central
membrane (CM), the end-caps, and the concentric  field cage cylinders.
The  central membrane  is operated  at  28 kV.   The end  caps are  at
ground.  The  field cage cylinders provide a  series of equi-potential
rings that divide the space between the central membrane and the anode
planes into  182 equally spaced segments.   One ring at  the center is
common to both  ends.  The central membrane is  attached to this ring.
The rings are  biased by resistor chains of  183 precision 2 M$\Omega$
resistors  which  provide  a  uniform  gradient  between  the  central
membrane and the grounded end caps.

The CM  cathode, a disk  with a central  hole to pass the  Inner Field
Cage (IFC), is made of 70  $\mu$m thick carbon loaded Kapton film with
a  surface resistance  of 230  $\Omega$ per  square.  The  membrane is
constructed from several pie  shape Kapton sections bonded with double
sided tape.  The membrane is secured under tension to an outer support
hoop  which is  mounted inside  the Outer  Field Cage  (OFC) cylinder.
There  is  no mechanical  coupling  to the  IFC  other  than a  single
electrical connection.  This design minimizes material and maintains a
good flat surface to within 0.5 mm.

Thirty six aluminum stripes have been  attached to each side of the CM
to provide  a low  work-function  material as the  target for  the TPC
laser    calibration    system\cite{laser,laser2}.    Electrons    are
photo-ejected  when ultraviolet  laser  photons hit  the stripes,  and
since the position  of the narrow stripes are  precisely measured, the
ejected electrons can be used for spatial calibration.

\begin{figure}[htb]
\includegraphics[width=14cm]{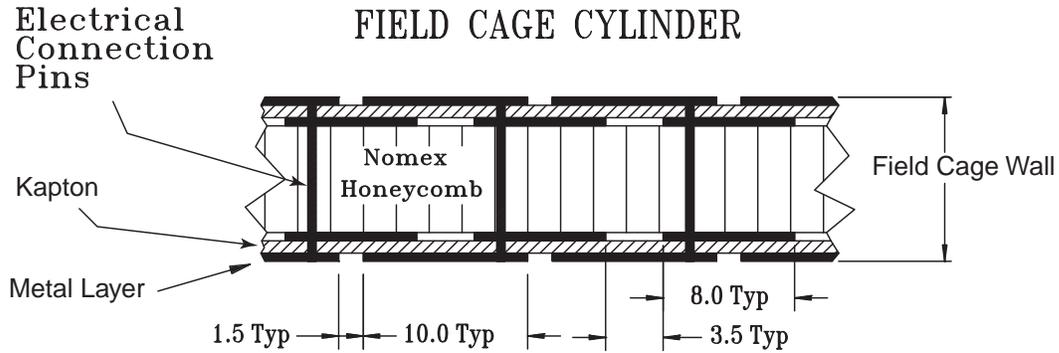}
\caption{A cutaway view of the inner field cage (IFC) showing the
construction and composition of the cylinder wall.  Dimensions are in mm.}
\label{cutaway}
\end{figure}

The  field  cage  cylinders  serve   the  dual  purpose  of  both  gas
containment and electric field  definition.  The mechanical design was
optimized  to reduce  mass, minimize  track distortions  from multiple
Coulomb  scattering,  and reduce  background  from secondary  particle
production.  Mechanically,  the walls of the low  mass self supporting
cylinders  are  effectively a  bonded  sandwich  of  two metal  layers
separated by NOMEX\cite{nomex} honeycomb  ( see Fig. \ref{cutaway} for
a cutaway view) .  The metal  layers are in fact flexible PC material,
Kapton,  with  metal on  both  sides.  The  metal  is  etched to  form
electrically separated 10 mm strips  separated by 1.5 mm.  The pattern
is  offset on  the  two sides  of  the kapton  so  that the  composite
structure  behaves mechanically  more like  a continuous  metal sheet.
The  1.5 mm break  is the  minimum required  to maintain  the required
voltage difference  between rings safely.  This  limits the dielectric
exposure in the drift  volume thus reducing stray, distorting electric
fields due to charge build  up on the dielectric surfaces.  Minimizing
the  break has  the  additional benefit  of  improving the  mechanical
strength.  Punch-through pins were  used to electrically  connect the
layers on the two sides of the sandwich.

The lay-up and bonding of the field cage sandwich was done on mandrels
constructed of wood covered with rigid foam which was turned to form a
good cylindrical surface.  Commercially available metal covered Kapton
is limited  in width  to $\approx$20  cm so the  lay-up was  done with
multiple etched  metal-Kapton sheets wrapped  around the circumference
of  the mandrel.   A laser  interferometer  optical tool  was used  to
correctly  position  the sheets  maintaining  the equi-potential  ring
alignment  to within  50  $\mu$m differentially  and  better than  500
$\mu$m,  overall.  The mandrels  were constructed  with a  double rope
layer under the  foam.  The ropes were unwound  to release the mandrel
from the field cage cylinder at completion of the lay-up.

A summary of  the TPC material thicknesses in  the tracking volume are
presented in Table \ref{tfieldcage}.  The design emphasis was to limit
material at the inner radius where multiple Coulomb scattering is most
important for accurate  tracking and accurate momentum reconstruction.
For this  reason aluminum  was used  in the IFC,  limiting it  to only
0.5\% radiation length ($X_0$).  To simplify the construction, and the
electrical connections,  copper was  used for the  OFC.  Consequently,
the OFC is significantly thicker, 1.3\% $X_0$, but still not much more
than  the detector  gas itself.   The  sandwich structure  of the  OFC
cylinder wall  is 10 mm  thick while the  IFC has a wall  thickness of
12.9 mm.

Nitrogen gas  or air insulation  was used to electrically  isolate the
field cage from the surrounding ground structures.  This design choice
requires more space than solid  insulators, but it has two significant
advantages.   One  advantage  is  to reduce  multiple  scattering  and
secondary particle production.  The  second advantage is the insulator
is not vulnerable  to permanent damage.  The gas  insulator design was
chosen  after  extensive  tests  showed  that the  field  cage  kapton
structures and  resistors could survive sparks with  the stored energy
of the full size field cage.  The  IFC gas insulation is air and it is
40 cm thick  without any detectors inside the IFC.  It  is 18 cm thick
with the  current suite  of inner detectors.   The OFC has  a nitrogen
layer  5.7 cm  thick isolating  it  from the  outer shell  of the  TPC
structure.   The field  cage surfaces  facing the  gas  insulators are
metallic  potential  graded  structures  which  are the  same  as  the
surfaces facing the  TPC drift volume.  In addition  to the mechanical
advantages of  a symmetric structure, this  design avoids uncontrolled
dielectric  surfaces where  charge migration  can lead  to  local high
fields and surface discharges in the gas insulator volume.

The outermost  shell of the TPC is  a structure that is  a sandwich of
material with  two aluminum skins separated by  an aluminum honeycomb.
The skins are  a multi-layer wraps of aluminum.   The construction was
done much  like the field  cage structures using the  same cylindrical
mandrel.   The  innermost  layer,  facing  the  OFC, is  electrically
isolated from the rest of the structure and it is used as a monitor of
possible corona  discharge across the gas insulator.   The outer shell
structure is  completely covered  by aluminum extrusion  support rails
bonded to  the surface.  The  support rails carry the  Central Trigger
Barrel (CTB) trays.  These extrusions have a central water channel for
holding the structure at  a fixed temperature.  This system intercepts
heat  from external  sources, the  CTB modules  and the  magnet coils,
which run at a temperature significantly higher than the TPC.  This is
just  one  part of  the  TPC  temperature  control system  which  also
provides cooling water for the TPC electronics on the end-caps.

\begin{table}[htb]

\vbox{
\begin{tabular}{|l|l|l|l|l|l|}
\hline
Structure &     Material &      Density($g/cm^3$) &     $X_0$
($g/cm^2$) &  Thickness (cm) & 
Thickness ($\% X_0$)\\
\hline
Insulating gas  & $N_2$ &    1.25E-03&  37.99 & 40    & 0.13 \\
TPC IFC         & Al    &       2.700&  24.01 & 0.004 & 0.04 \\
TPC IFC         & Kapton&       1.420&  40.30 & 0.015 & 0.05 \\
TPC IFC         & NOMEX &       0.064&  40    & 1.27  & 0.20 \\
TPC IFC         &Adhesive &     1.20 &  40    & 0.08  & 0.23 \\
\hline
IFC Total (w/gas)&        &          &        &       & 0.65 \\
\hline
\end{tabular}

\vspace{0.1cm}

\begin{tabular}{|l|l|l|l|l|l|}
\hline
Structure &     Material &      Density($g/cm^3$) &     $X_0$
($g/cm^2$) &  Thickness (cm) & 
Thickness ($\% X_0$)\\
\hline
TPC gas  &      P10 &   1.56E-03 & 20.04 &      150.00  & 1.17 \\
TPC OFC  &      Cu  &   8.96     & 12.86 &      0.013   & 0.91 \\
TPC OFC  &  Kapton  &   1.420    & 40.30 &      0.015   & 0.05 \\
TPC OFC  &   NOMEX  &   0.064    & 40    &      0.953   & 0.15 \\
OFC      & Adhesive &   1.20     & 40    &      0.05    & 0.15 \\
\hline
OFC Total (w/gas)& & & & & 2.43 \\
\hline
\end{tabular}
}
\vspace{0.2cm}
\caption[]{Material thickness for the inner (IFC) and outer (OFC)
electrostatic field cages\cite{est} .}
\label{tfieldcage}
\end{table}

\section{The TPC End-caps with the Anodes and Pad Planes}

The end-cap  readout planes of STAR closely match  the designs used in
other TPCs such as PEP4, ALEPH, EOS and NA49 but with some refinements
to accommodate  the high  track density at  RHIC and some  other minor
modifications to  improve reliability and  simplify construction.  The
readout  planes, MWPC  chambers with  pad readout,  are  modular units
mounted on  aluminum support wheels.   The readout modules, or sectors,
are arranged  as on a  clock with 12  sectors around the  circle.  The
modular  design with manageable  size sectors  simplifies construction
and  maintenance.  The  sectors are  installed  on the  inside of  the
spoked support  wheel so that there  are only 3 mm  spaces between the
sectors.  This reduces  the dead area between the  chambers, but it is
not hermetic  like the more complicated  ALEPH TPC design\cite{BandR}.
The simpler  non-hermetic design was  chosen since it is  adequate for
the physics in the STAR experiment.

\begin{figure}[htb]
\includegraphics[width=14cm]{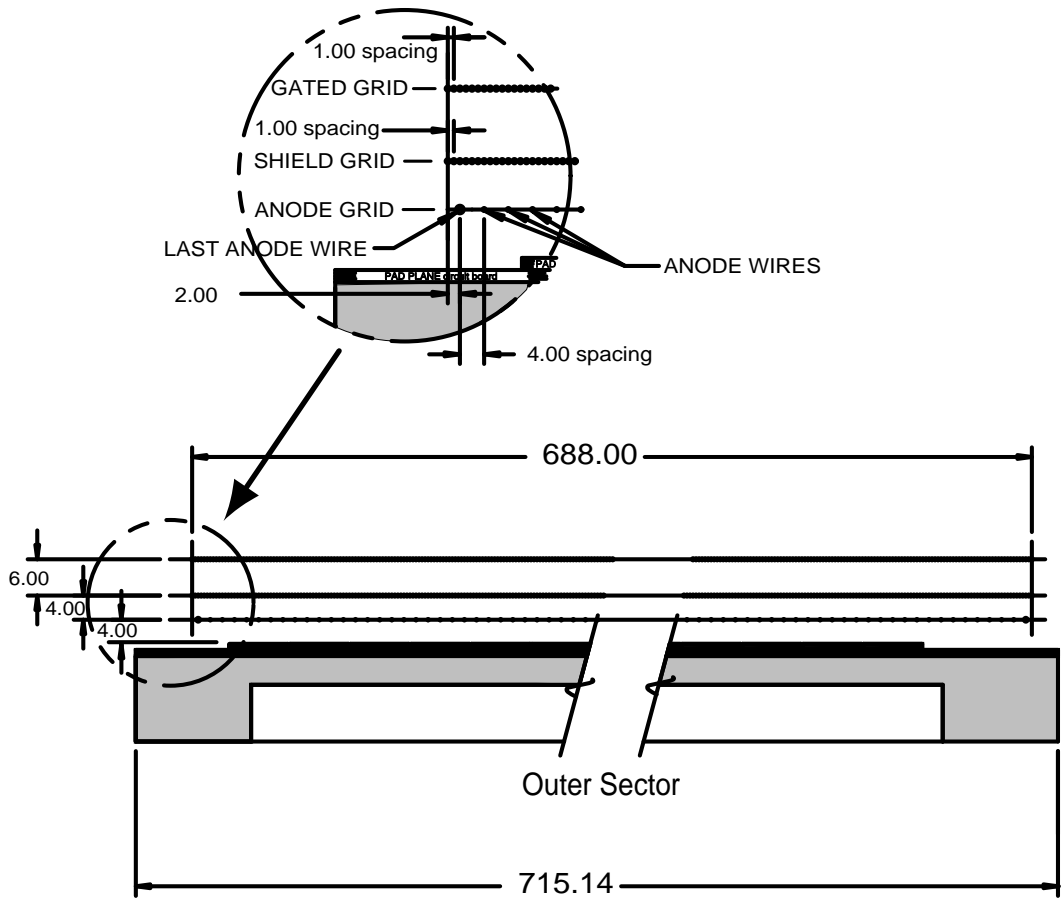}
\caption{A cutaway view  of an outer sub-sector pad  plane. The cut is
taken along  a radial  line from the  center of  the TPC to  the outer
field cage  so the center  of the detector  is towards the  right hand
side of the figure.  The figures  shows the spacing of the anode wires
relative to the pad plane, the ground shield grid, and the gated grid.
The  bubble diagram shows  additional detail  about the  wire spacing.
The inner sub-sector pad plane  has the same layout except the spacing
around the anode  plane is 2 mm  instead of the 4 mm  shown here.  All
dimensions are in millimeters.}
\label{cutaway2}
\end{figure}

\begin{figure}[htb]
\includegraphics[width=14cm]{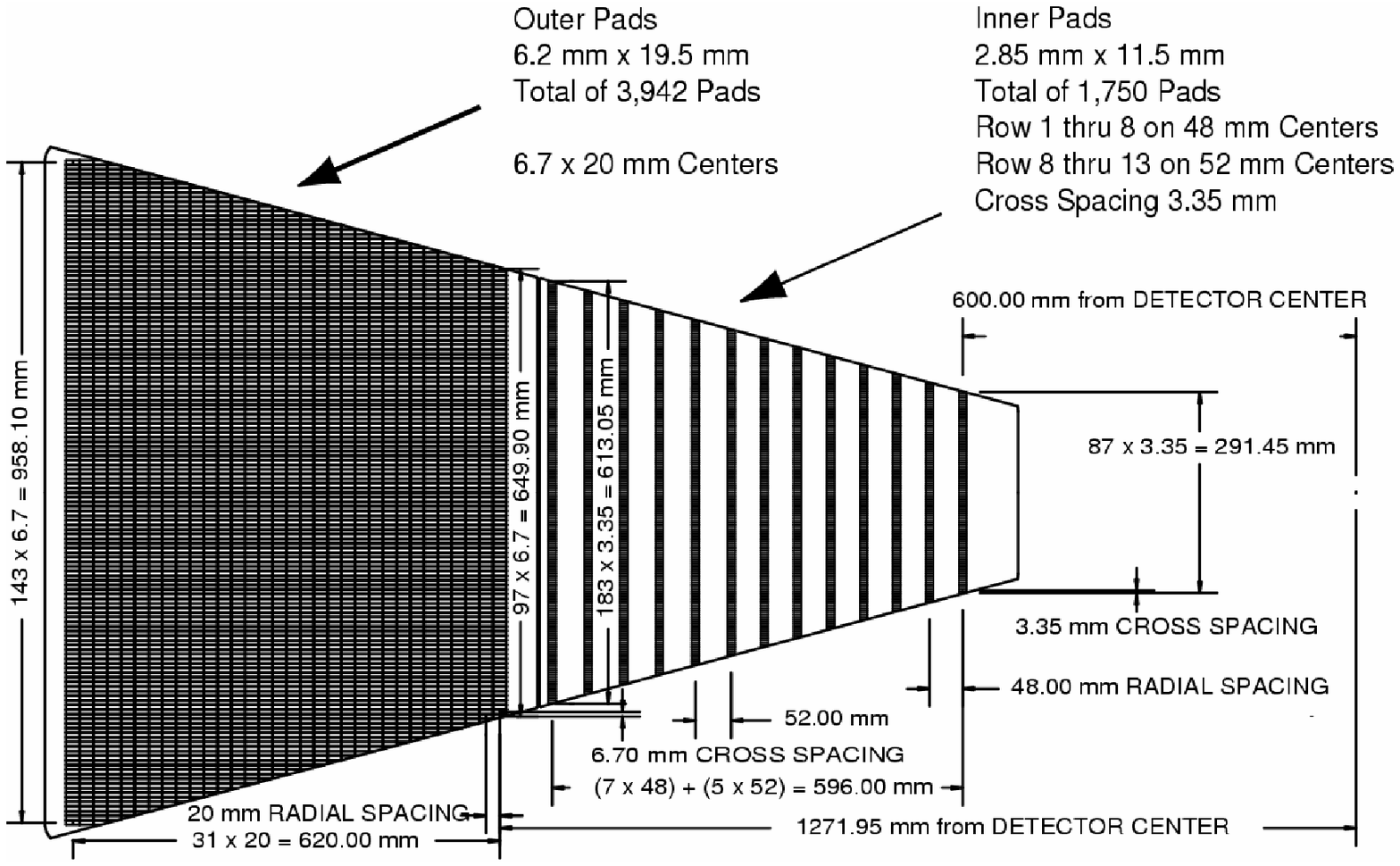}
\caption{The anode pad plane with one full sector shown. The inner
  sub-sector is on the right and  it has small pads arranged in widely
  spaced rows.  The outer sub-sector is  on the left and it is densely
  packed with larger pads.}
\label{anodeplane}
\end{figure}

The chambers  consist of four components;  a pad plane  and three wire
planes (see  Fig. \ref{cutaway2}). The  amplification/readout layer is
composed of the  anode wire plane of small, 20  $\mu$m, wires with the
pad plane  on one side  and the ground  wire plane on the  other.  The
third  wire plane  is a  gating grid  which will  be  discussed later.
Before  addressing the  details of  the amplification  region,  a word
about  the  chosen wire  direction.   The  direction  is set  to  best
determine  the momentum  of  the highest  transverse momentum  ($p_T$)
particles whose tracks are nearly straight radial lines emanating from
the interaction  point (the  momentum of low  $p_T$ particles  is well
determined without  special consideration).   The sagitta of  the high
$p_T$  tracks is  accurately  determined by  setting  the anode  wires
roughly perpendicular  to the straight radial  tracks because position
resolution  is best along  the direction  of the  anode wire.   In the
other direction, the resolution is limited by the quantized spacing of
the  wires  (4  mm  between  anode  wires).   The  dimensions  of  the
rectangular  pads are  likewise optimized  to give  the  best position
resolution perpendicular  to the stiff  tracks.  The width of  the pad
along the wire  direction is chosen such that  the induced charge from
an avalanche point on the wire  shares most of it's signal with only 3
pads.   This is  to  say that  the optimum  pad  width is  set by  the
distance  from the  anode wire  to the  pad plane.   Concentrating the
avalanche  signal on  3 pads  gives the  best  centroid reconstruction
using either a  3-point gaussian fit or a  weighted mean.  Accuracy of
the centroid determination depends  on signal-to-noise ratio and track
angle,  but  it  is typically  better  than  20\%  of the  narrow  pad
dimension.  There  are additional  tradeoffs dictating details  of the
pads' dimensions which will be discussed further in connection with our
choice  of two  different sectors  designs, one  design for  the inner
radius where track density is  highest and another design covering the
outer radius region.   Details of the two sector  designs can be found
in Table \ref{innerouter} and Figure \ref{anodeplane}.

The outer radius sub-sectors  have continuous pad coverage to optimize
the  $dE/dx$ resolution  (ie. no  space  between pad  rows).  This  is
optimal because the full track ionization signal is collected and more
ionization  electrons improve statistics  on the  $dE/dx$ measurement.
Another modest  advantage of  full pad coverage  is an  improvement in
tracking  resolution due  to  anti-correlation of  errors between  pad
rows.  There is an error in position determination for tracks crossing
a pad  row at an angle  due to granularity in  the ionization process
(Landau fluctuations).   If large clusters of ionization  occur at the
edge of the pad row they pull the measured centroid away from the true
track  center.  But,  there is  a partially  correcting effect  in the
adjacent pad row. The large clusters at the edge also induce signal on
the adjacent  pad row  producing an oppositely  directed error  in the
measured  position in this  adjacent row.   This effective  cross talk
across pad rows, while helpful  for tracking precision, causes a small
reduction in $dE/dx$ resolution.

\begin{table}[htb]
\begin{tabular}{|l|l|l|l|}
\hline
Item & Inner Subsector & Outer Subsector & Comment \\
\hline
Pad Size        & 2.85 mm x 11.5 mm &   6.20 mm x 19.5 mm & \\
Isolation Gap between pads & 0.5 mm & 0.5 mm & \\
Pad Rows        & 13 (\#1-\#13)    & 32 (\#14-\#45) &  \\
Number of Pads  &  1,750 & 3,942 & 5,692 total \\
Anode Wire to Pad Plane Spacing  &      2 mm &  4 mm & \\
Anode Voltage & 1,170 V & 1,390 V &  20:1 signal:noise  \\
Anode Gas Gain &    3,770 & 1,230 & \\ 
\hline
\end{tabular}
\vspace{0.2cm}
\caption{Comparison of the Inner and Outer subsector geometries.}
\label{innerouter}
\end{table}

On the outer radius sub-sectors the pads are arranged on a rectangular
grid with a pitch of 6.7  mm along the wires and 20.0 mm perpendicular
to the wires.  The grid is phased  with the anode wires so that a wire
lies over  the center of  the pads.  There  is a 0.5 mm  isolation gap
between pads.   The 6.7  mm pitch  and the 4  mm distance  between the
anode wire plane is consistent  with the transverse diffusion width of
the electron  cloud for tracks that  drift the full  2 meter distance.
More explicitly,  with a 4 mm  separation between pad  plane and anode
plane the width  of the induced surface charge  from a point avalanche
is the  same as the diffusion width.   The pad pitch of  6.7 mm places
most of the  signal on 3 pads which  gives good centroid determination
at minimum gas gain.  This matching gives good signal to noise without
serious compromise to two-track  resolution.  The pad size in the long
direction  (  20.0  mm  pitch)  was  driven  by  available  electronic
packaging  density  and  funding,   plus  the  match  to  longitudinal
diffusion.  The z  projection of 20.0 mm on $\eta$  = 1 tracks matches
the longitudinal diffusion spread in  z for $\eta$ = 0 tracks drifting
the full two meters.
 
The inner sub  sectors are in the region of  highest track density and
thus  are optimized  for good  two-hit resolution.   This  design uses
smaller pads which are 3.35 mm by 12 mm pitch.  The pad plane to anode
wire  spacing is  reduced accordingly  to 2  mm to  match  the induced
signal width to ~3 pads.   The reduction of the induced surface charge
width to  less than  the electron cloud  diffusion width  improves two
track   resolution  a   small  amount   for  stiff   tracks  $\approx$
perpendicular  to  the  pad  rows  at  $\eta$  $\approx$0.   The  main
improvement in  two track resolution,  however, is due to  shorter pad
length (12 mm instead of 20 mm).  This is important for lower momentum
tracks which  cross the pad row  at angles far  from perpendicular and
for  tracks  with  large  dip  angle.  The  short  pads  give  shorter
projective widths in the  r-$\phi$ direction (the direction along the
pad row), and the z direction  (the drift direction)  for these angled
tracks.  The compromise inherent in the inner radius sub-sector design
with  smaller  pads  is  the  use  of separate  pad  rows  instead  of
continuous  pad coverage.   This constraint  imposed by  the available
packing density of  the front end electronics channels  means that the
inner  sector  does  not  contribute significantly  to  improving  the
$dE/dx$  resolution.   The inner  sector  only  serves  to extend  the
position measurements  along the track  to small radii  thus improving
the  momentum  resolution  and  the  matching to  the  inner  tracking
detectors.  An additional benefit is detection of particles with lower
momentum.

The design  choices, pad sizes, and  wire-to-pad spacing,  for the two
pad  plane  sector geometries  were  verified  through simulation  and
testing with computer models  \cite{CDR,CDRU}, but none of the desired
attributes:  $dE/dx$  resolution, momentum  resolution  and two  track
resolution show a dramatic  dependence on the design parameters.  This
is in part  due to the large variation in track  qualities such as dip
angle, drift distance,  and crossing angle.  While it  is not possible
with a TPC to focus the  design on a particular condition and optimize
performance, a lot is  gained through over-sampling and averaging.  In
addition to simulations, prototype pad chambers were built and studied
to verify charge-coupling parameters and to test stability at elevated
voltages \cite{betts}.

The anode wire plane has one  design feature that is different than in
other TPCs.  It is  a single plane of 20 $\mu$m wires  on a 4 mm pitch
without intervening field wires.  The elimination of intervening field
wires  improves  wire  chamber  stability and  essentially  eliminates
initial voltage conditioning time.  This is because in the traditional
design  both the  field wire  and the  anode wires  are captured  in a
single epoxy  bead.  The large  potential difference on the  field and
anode wires places significant  demands on the insulating condition of
the epoxy surface. The surface is much less of a problem in our design
where the epoxy  bead supports only one potential.   This wire chamber
design  requires a  slightly  higher  voltage on  the  anode wires  to
achieve  the same electric  field at  the anode  wire surface  (i.e. a
higher  voltage to  achieve  the same  gas  gain) but  this  is not  a
limitation on  stability.  Another small  advantage in this  design is
that we can  operate the chambers at a lower gas  gain (35\% lower for
the inner sector) \cite{betts} since with this design the readout pads
pick-up a  larger fraction of  the total avalanche signal.   Like other
TPCs, the  edge wires on the  anode wire plane are  larger diameter to
prevent the excess gain that would otherwise develop on the last wire.

Most   of  the   anode  wires   are  equipped   with   amplifiers  and
discriminators that are  used in the trigger to  detect tracks passing
through  the  end  cap.   The  discriminators are  active  before  the
electrons drift in from tracks in the drift volume.

Another special feature of the anode  plane is a larger than normal (1
nF) capacitor to ground on each wire.  This reduces the negative cross
talk  that is  always induced  on the  pads under  a wire  whenever an
avalanche  generates charge  anywhere  along the  wire.  The  negative
cross talk  comes from  capacitive coupling between  the wire  and the
pad.  The AC component of  the avalanche charge on a wire capacitively
couples to  the pads proportionally as $C_p/C_{total}$  where $C_p$ is
the pad-to-wire  capacitance and $C_{total}$ is  the total capacitance
of the wire  to ground.  In the high track density  at RHIC, there can
be multiple  avalanches on a  wire at any  time so it is  important to
minimize  this source  of cross-talk  and noise.   The 1  nF grounding
capacitor is a compromise between cross talk reduction and wire damage
risk.  Our  tests showed that  the stored energy in  larger capacitors
can damage the wire in the event of a spark.

The  gas gain,  controlled by  the anode  wire voltage,  has  been set
independently for  the two sector types  to maintain a  20:1 signal to
noise for pads intercepting the center of tracks that have drifted the
full   2  meters.    This   choice  provides   minimum  gain   without
significantly impacting  the reconstructed position  resolution due to
electronic  noise.  The  effective  gas gain  needed  to achieve  this
signal to noise is 3,770 for  the inner sector and 1,230 for the outer
sector.  As  discussed in detail  in Ref. \cite{readout}  the required
gas gain  depends on diffusion size  of the electron  drift cloud, pad
dimensions,    amplifier    shaping    time,   the    avalanche-to-pad
charge-coupling fractions and the electronic noise which for our front
end electronics is $\approx$1000 electrons rms.

The ground  grid plane of 75  $\mu$m wires completes  the sector MWPC.
The primary  purpose of the ground  grid is to terminate  the field in
the avalanche region and provide additional rf shielding for the pads.
This  grid can also  be pulsed  to calibrate  the pad  electronics.  A
resistive divider at the grid provides 50 $\Omega$ termination for the
grid and and 50 $\Omega$ termination for the pulser driver.

The outermost  wire plane on the  sector structure is  the gating grid
located 6 mm from the ground  grid.  This grid is a shutter to control
entry of electrons  from the TPC drift volume into  the MWPC.  It also
blocks positive ions produced in  the MWPC, keeping them from entering
the drift volume where they would distort the drift field.  The gating
grid plane  can have  different voltages on  every other wire.   It is
transparent  to  the drift  of  electrons  while  the event  is  being
recorded and closed the rest of the time.  The grid is `open' when all
of the wires are biased to  the same potential (typically 110 V).  The
grid  is `closed'  when the  voltages alternate  $\pm$ 75  V  from the
nominal value.   The positive ions are  too slow to  escape during the
open  period and  get captured  during  the closed  period.  The  STAR
gating  grid  design  is  standard.   Its  performance  is  very  well
described by the usual equations \cite{BandR}.  The gating grid driver
has  been designed  to open  and  settle rapidly  (100 V  in 200  ns).
Delays  in opening  the  grid shorten  the  active volume  of the  TPC
because electrons that drift into  the grid prior to opening are lost.
The combined  delay of  trigger plus the  opening time for  the gating
grid is  2.1 $\mu$s. This means  that the useful length  of the active
volume is  12 cm less  than the physical  length of 210 cm.   To limit
initial data corruption at the opening of the gate, the plus and minus
grid driving voltages are well matched in time and amplitude to nearly
cancel the induced signal on the pads.

The  gating  grid establishes  the  boundary  conditions defining  the
electric field  in the TPC drift volume  at the ends of  the TPC.  For
this reason the gating wire  planes on the inner and outer sub-sectors
are aligned on  a plane to preserve the uniform  drift field.  For the
same reason the potential on the gating grid planes must be matched to
the potential on  the field cage cylinders at  the intersection point.
Aligning the gating grid plane  separates the anode wire planes of the
two sector types by 2 mm.  The difference in drifting electron arrival
time for  the two  cases is taken  into account  in the time-to-space
position calibration.  The time difference is the result of both the 2
mm offset  and the  different field strengths  in the vicinity  of the
anode wires for  the two sector types.  The  electron drift times near
the  anode plane  was both  measured and  studied with  MAGBOLTZ.  The
field is nearly uniform and constant from the CM to within 2 mm of the
gating grid.  We simulated the drift of ionization from 2 mm above the
gating grid to the anode  wires to estimate the difference between the
inner and  outer sub-sector  drift times.  These  MAGBOLTZ simulations
find  that the drift  from the  CM to  the outer  sub-sectors requires
0.083  $\mu$s  longer than  from  the  CM  to the  inner  sub-sectors.
Measurement shows  a slightly longer average time  difference of 0.087
$\mu$s.

The  construction of  the  sectors followed  techniques developed  for
earlier  TPCs.  The  pad planes  are constructed  of  bromine-free G10
printed  circuit  board  material  bonded to  a  single-piece  backing
structure machined from solid  aluminum plate. Specialized tooling was
developed  so that  close tolerances  could be  achieved  with minimum
setup time.  Pad plane flatness  was assured by vacuum locking the pad
plane  to a flat  granite work  surface while  the aluminum  backer is
bonded with  epoxy to the pad  plane.  Wire placement is  held to high
tolerance with fixed combs on  granite work tables during the assembly
step  of capturing  the wires  in epoxy  beads on  the  sector backer.
Mechanical    details   of    the   wires    are   given    in   Table
\ref{wirecomparison}.  The  final wire-placement error is  less than 7
$\mu$m.  Pad location along the plane is controlled to better than 100
microns.   The sectors  were qualified  with over-voltage  testing and
gas-gain uniformity measurements with an $^{55}$Fe source.

\section{Drift Gas}

P10 (90\%  Argon + 10\% Methane) is  the working gas in  the TPC.  The
gas system (discussed  in detail in \cite{gas}) circulates  the gas in
the  TPC and  maintains purity,  reducing electro  negative impurities
such as  oxygen and water  which capture drifting electrons.   To keep
the electron absorption to a few percent, the oxygen is held below 100
parts per million and water less than 10 parts per million.

All materials  used in  the TPC construction  that are exposed  to the
drift  gas   were  tested   for  out-gassing  of   electron  capturing
contaminants.   This  was done  with  a  chamber  designed to  measure
electron attenuation by drifting electrons  through a 1 meter long gas
sample.

The transverse diffusion\cite{Magboltz} in P10 is $230 \mu m/\sqrt{\rm
cm}$ at  0.5 T  or about $\sigma_T$  = 3.3  mm after drifting  210 cm.
This sets  the scale for  the wire chamber  readout system in  the X,Y
plane.   Similarly,  the  longitudinal   diffusion  of  a  cluster  of
electrons that drifts  the full length of the TPC  is $\sigma_L$ = 5.2
mm.  At a drift velocity of 5.45 cm/$\mu$s, the longitudinal diffusion
width is  equal to a spread  in the drift  time of about 230  ns FWHM.
This diffusion width sets the scale for the resolution of the tracking
system in  the drift  direction and we  have chosen the  front-end pad
amplifier shaping  time and the electronic  sampling time accordingly.
The shaping  time is 180 ns  FWHM and the electronic  sampling time is
9.4 MHz.

\begin{table}[htb]
\begin{tabular}{|l|l|l|l|l|}
\hline
Wire & Diameter & Pitch & Composition & Tension \\
\hline
Anodes & $20\ \mu$m & 4 mm & Au-plated W & 0.50 N \\
Anodes - Last wire & 125 $\mu$m & 4 mm & Au-plated Be-Cu & 0.50 N \\
Ground Plane & $75\ \mu$m  & 1 mm & Au-plated Be-Cu & 1.20 N \\
Gating Grid & $75\ \mu$m  & 1 mm & Au-plated Be-Cu & 1.20 N \\
\hline
\end{tabular}
\vspace{0.2cm}
\caption{Properties of the wires in the readout chambers.}
\label{wirecomparison}
\end{table}

\section{Performance of the TPC}

This section will discuss the  TPC performance using data taken in the
RHIC beam in the 2000/2001  run cycle. The TPC performance with cosmic
rays without magnetic field has been previously presented\cite{stest}.
In 2000, the  magnetic field was 0.25 T; in 2001  the field was raised
to 0.5  T.  The TPC performance  is strongly affected  by the magnetic
field because, for example,  the transverse diffusion of the electrons
that drift through the gas is smaller in higher fields.

The track of an infinite-momentum  particle passing through the TPC at
mid-rapidity is  sampled by 45 pad  rows, but a  finite momentum track
may not cross  all 45 rows.  It depends on the  radius of curvature of
the  track,  the  track  pseudorapidity,  fiducial  cuts  near  sector
boundaries, and other details  about the particle's trajectory.  While
the  wire chambers  are sensitive  to  almost 100\%  of the  secondary
electrons arriving at the  end-cap, the overall tracking efficiency is
lower (80-90\%) due to the fiducial cuts, track merging, and to lesser
extent bad  pads and dead channels.   There are at most  a few percent
dead channels in any one run cycle.

The  track  of   a  primary  particle  passing  through   the  TPC  is
reconstructed  by finding  ionization clusters  along the  track.  The
clusters are found  separately in $x,y$ and in  $z$ space.  (The local
$x$ axis  is along the direction of  the pad row, while  the local $y$
axis  extends from  the beamline  outward through  the middle  of, and
perpendicular  to, the pad  rows.  The  $z$ axis  lies along  the beam
line.)   For  example,  the  $x$-position  cluster  finder  looks  for
ionization on  adjacent pads,  within a pad  row, but  with comparable
drift times.   And, for simple clusters,  the energy from  all pads is
summed to give the total ionization in the cluster.  If two tracks are
too  close  together, the  ionization  clusters  will overlap.   These
complex clusters  are split  using an algorithm  that looks  for peaks
with a valley between them  and then the ionization is divided between
the two tracks.  These merged  clusters are used only for tracking and
not  for  $dE/dx$ determination  because  of  the  uncertainty in  the
partitioning between the tracks.  In  central Au-Au events at 200 GeV,
about 30\% of the clusters are overlapping.

\subsection{Reconstruction of the $x$, $y$ position}

The $x$ and $y$ coordinates of  a cluster are determined by the charge
measured  on adjacent pads  in a  single pad  row.  Assuming  that the
signal distribution  on the pads (pad response  function) is Gaussian,
the local $x$ is given by a  fit, where $h_1$, $h_2$ and $h_3$ are the
amplitudes on 3 adjacent pads, with pad $h_2$ centered at $y=0$:

\begin{equation}
x = {\sigma^2\over 2w}\ln{\big({h_3\over h_1}\big)}
\end{equation}
where the width of the signal, $\sigma$, is given by
\begin{equation}
\sigma^2={w^2 \over \ln{(h_2^2/h_1h_3)}}
\end{equation}

and $w$ is the pad width.  The position uncertainty due to electronics
noise may be fairly easily computed in this approach:

\begin{equation}
\Delta x = {\Delta h\over h_c} {\sigma^2\over 2 w}
\sqrt{\big(1-{2x\over w}\big)^2\exp{\big({-(x+w)^2\over\sigma^2}\big)} +
{16x^2\over w^2} \exp{\big(-{x^2\over w^2}\big)} +
\big(1+{2x\over w}\big)^2\exp{\big({-(x-w)^2\over\sigma^2}\big)}}
\end{equation}

Here $\Delta h$  is the noise, $h_c$ is the  signal amplitude under a
centerd pad ($h_c=0$),  and the three terms in  the root correspond to
the the errors on $h_1$, $h_2$, and $h_3$ respectively.  For $\Delta h
< 0.05  h$ (a 20:1 signal  to noise ratio), the  noise contribution is
small.   The total  signal  is summed  over  all above-threshold  time
buckets.
This   equation   is   slightly   different  from   the   results   in
Ref.  \cite{lynch}  because it  includes  the  error  in the  $\sigma$
determination.

The Gaussian approximation has  some short-comings. First, it does not
exactly  match onto  the tails  the true  pad response  function which
introduces  an $x-$depend\-ent  bias of  a few  hundred  $\mu$m.  More
importantly,  the  algorithm deteriorates  at  large crossing  angles.
When  a  track  crosses the  pad  row  at  large angles,  it  deposits
ionization  on many pads  and any  3 adjacent  pads will  have similar
amplitude signals.  In this case, a weighted mean algorithm, using all
of the pads above a certain threshold is much more effective.

\begin{figure}[htb]
\includegraphics[width=14cm]{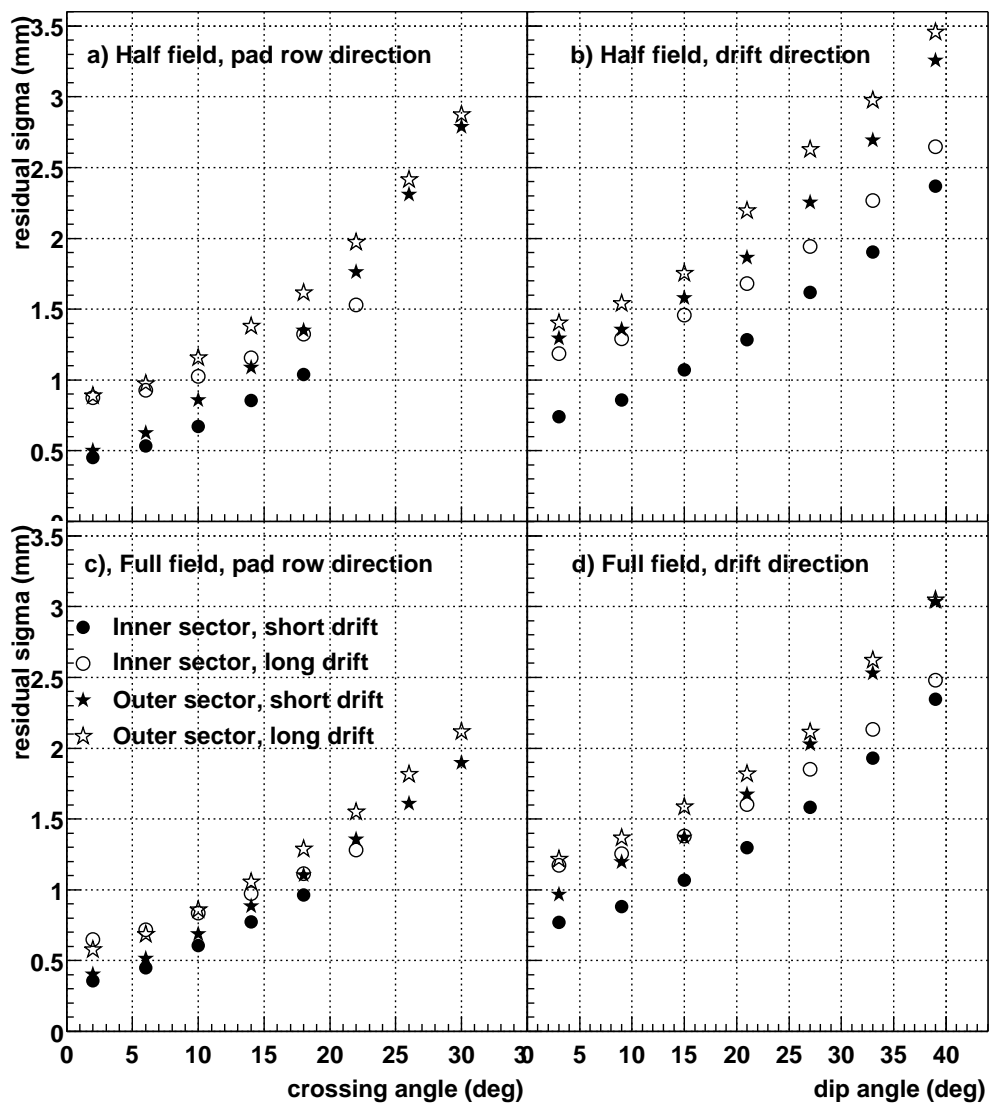}
\caption{Position resolution across the pad rows and along the $z$
  axis  of  the TPC.  
The  crossing angle  is  the  angle between  the
  particle momentum  and the pad row  direction. The dip  angle is the
  angle between the particle momentum and the drift direction, $\theta
  = \cos^{-1}{(p_z/p)}$.}
\label{position}
\end{figure}

Figures   \ref{position}a  and   \ref{position}c  show   the  position
resolution along the  pad rows (local $x$) for  both field settings of
the  magnet. The  sigma  is extracted  by  fitting a  Gaussian to  the
residual distribution, i.e. the  distance between the hit position and
the track extrapolation.

\subsection{Reconstruction of the $z$ position in the TPC}

The  $z$  coordinate  of a  point  inside  the  TPC is  determined  by
measuring the time  of drift of a cluster  of secondary electrons from
the point  of origin to the anodes  on the endcap and  dividing by the
average drift velocity. The arrival  time of the cluster is calculated
by measuring the time of  arrival of the electrons in ``time buckets''
and weighting  the average by the  amount of charge  collected in each
bucket. (Each time bucket is  approximately 100 nsec long.) The signal
from a  typical cluster covers  several time buckets because  of three
phenomena: the  longitudinal diffusion of the  drifting electrons, the
shaping of the  signal by the preamplifier electronics,  and the track
dip angle.  The  preamplifier shaping time is chosen  to correspond to
the size  of the electron  cloud for particles drifting  and diffusing
the entire length of the  TPC\cite{FEE}.  This setting smooths out the
random  fluctuations of  the average  cluster positions  introduced by
statistics  and  diffusion.    The  amplifier  also  has  cancellation
circuitry  to   remove  the   long  current  tail   characteristic  of
MWPCs\cite{SAS}.

The length  of the  signal reaching  a pad depends  on the  dip angle,
$\theta$, which  is the  angle between the  particle momentum  and the
drift direction.  The ionization  electrons are spread over a distance
$d$ along the  beam axis, with $d  = L / \tan(\theta)$ and  $L$ is the
length of the  pad.

The drift velocity  for the electrons in the gas must  be known with a
precision  of 0.1\%  in  order to  convert  the measured  time into  a
position with sufficient accuracy.  But the drift velocity will change
with atmospheric pressure because the  TPC is regulated and fixed at 2
mbar above atmospheric pressure.  Velocity changes can also occur from
small  changes in  gas compositon.  We  minimize the  effect of  these
variations  in two ways.   First, we  set the  cathode voltage  so the
electric  field  in the  TPC  corresponds to  the  peak  in the  drift
velocity curve  (i.e.  velocity vs.  electric field  / pressure).  The
peak is broad and flat and  small pressure changes do not have a large
effect  on the drift  velocity at  the peak.   Second, we  measure the
drift velocity  independently every few hours  using artificial tracks
created  by lasers  beams\cite{laser,laser2}.   Table \ref{majorparam}
gives the typical drift velocities and cathode potentials.

The conversion from time to position also depends on the timing of the
first  time bucket  with respect  to  the collision  time.  This  time
offset  has several  origins: trigger  delay,  the time  spent by  the
electron drifting from the gating grid to the anode wires, and shaping
of the signal in the front end electronics. The delay is constant over
the full volume  of the TPC and so the timing  offset can be adjusted,
together with  the drift  velocity, by reconstructing  the interaction
vertex using data from one side  of the TPC only and later matching it
to the vertex  found with data from the other side  of the TPC.  Local
variations of  the time offset  can appear due to  differences between
different electronic channels and  differences in geometry between the
inner and  outer sector pad  planes.  These electronic  variations are
measured  and corrected  for by  applying  a calibrated  pulse on  the
ground  plane. Fluctuations  on  the  order of  0.2  time buckets  are
observed between different channels.

Figures   \ref{position}b  and   \ref{position}d  show   the  position
resolution along  the $z$ axis of  the TPC in 0.25T  and 0.5T magnetic
fields, respectively. The resolution is best for short drift distances
and small  dip angles.  The  position resolution depends on  the drift
distance but the dependence is  weak because of the large shaping time
in  the  electronics, which  when  multiplied  by  the drift  velocity
($\approx 1$  cm), is comparable  to or greater than  the longitudinal
diffusion width  ($\approx 0.5$ cm).  The position resolution  for the
two magnetic field settings  is similar.  The resolution deteriorates,
however, with increasing dip angle because the length of path received
by a  pad is greater than  the shaping time of  the electronics (times
drift  velocity) and  the ionization  fluctuations along  the particle
path are not fully integrated out of the problem.

\subsection{Distortions}

The position of a secondary electron at the pad plane can be distorted
by  non-uniformities  and global  misalignments  in  the electric  and
magnetic fields of  the TPC.  The non-uniformities in  the fields lead
to a  non-uniform drift of the  electrons from the point  of origin to
the pad plane.  In the STAR TPC, the electric  and magnetic fields are
parallel and nearly uniform in  $r$ and $z$. The deviations from these
ideal conditions are small and  a typical distortion along the pad row
is $\leq$ 1 mm before applying corrections.

Millimeter-scale distortions  in the direction transverse  to the path
of  a  particle,  however,  are  important  because  they  affect  the
transverse  momentum determination  for particles  at high  $p_T$.  In
order  to understand  these  distortions, and  correct  for them,  the
magnetic field was carefully mapped  with Hall probes and an NMR probe
before the TPC  was installed in the magnet\cite{magnet}.   It was not
possible to measure the electric fields and so we calculated them from
the known geometry of the TPC.   With the fields known, we correct the
hit positions  along the pad  rows using the distortion  equations for
nearly parallel electric and magnetic fields\cite{BandR}.

\begin{equation}
\delta_x = \int {-\omega\tau{B}_y + \omega^2\tau^2{B}_x \over
(1+\omega^2\tau^2){B}_z} dz +
\int {{E}_x + \omega\tau{E}_y \over 
(1+\omega^2\tau^2){E}_z} dz
\end{equation}

\begin{equation}
\delta_y = \int {\omega\tau{B}_x + \omega^2\tau^2{B}_y \over
(1+\omega^2\tau^2){B}_z} dz +
\int {{E}_y - \omega\tau{E}_x \over 
(1+\omega^2\tau^2){E}_z} dz
\end{equation}

where $\delta_x$ is the distortion in the $x$ direction, $\vec{E}$ and
$\vec{B}$ are the electric and magnetic fields, $\omega$ is the signed
cyclotron  frequency, and  $\tau$ is  the characteristic  time between
collisions as the electron diffuses through the gas.

These  are precisely  the equations  in Blum  and Rolandi\cite{BandR},
except that they are valid  for any $\vec{E}$ field or $\vec{B}$ field
configuration while  the equations in  Blum and Rolandi are  {\bf not}
valid for all orientations  of $\vec{E}$ and $\vec{B}$.  Our equations
differ from  Blum and Rolandi  in the definition of  $\omega\tau$.  In
Blum and Rolandi, $\omega\tau$ is always positive.  Here, $\omega\tau$
is signed, with  the sign depending on the  directions of ${B}_z$,
${E}_z$ and the drift velocity ${u}_z$:

\begin{equation}
\omega\tau = 
k {{u}_z (cm/\mu s) \over {E}_z(V/cm)} {B}_z (T)
\end{equation}

where  $k$  is  a  constant.   The negative  charge  of  the  drifting
electrons is  included in  the sign of  $u_z$.  For example,  the STAR
electric  field  always  points   towards  the  central  membrane  and
electrons always drift  away from it, while $B_z$  can point in either
direction.  Here, $k\approx 100$ and it depends on microscopic physics
that is not  represented in equations 4 and 5.   For precise work, $k$
must      be       determined      by      measuring      $\omega\tau$
directly\cite{BandR,amend}.  In  STAR, $k=110$ and  so $|\omega\tau| =
1.15$  at 0.25  T, rising  to  $|\omega\tau| =  2.30$ at  0.5 T.   The
magnitude   of  the   distortion  corrections   are  given   in  Table
\ref{tdistort}.

\begin{figure}[htb]
\includegraphics[width=14cm]{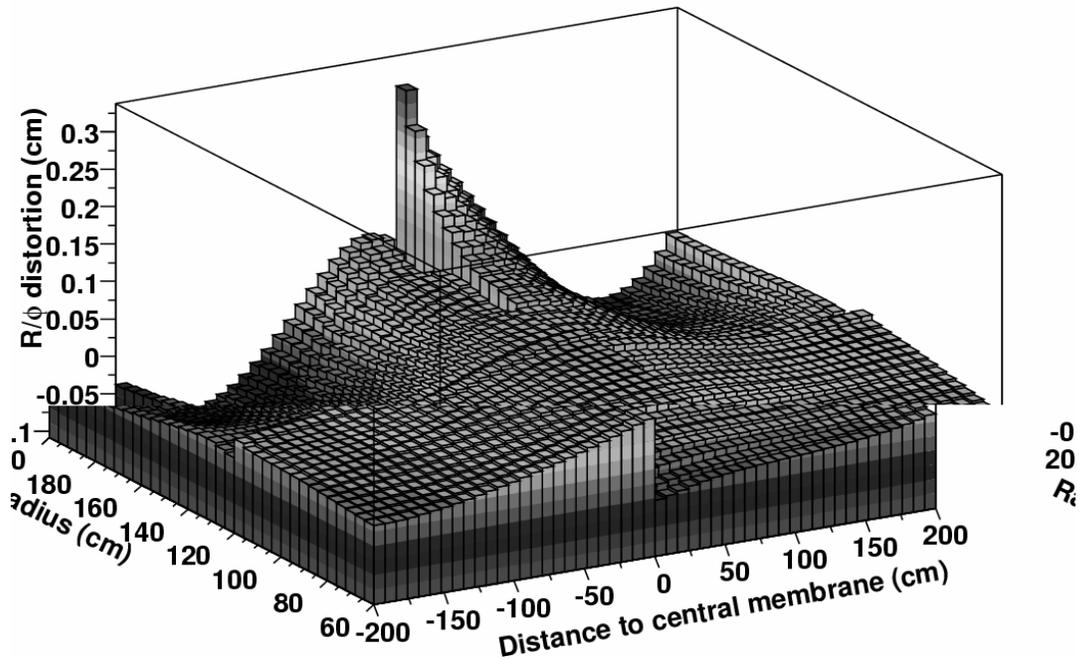}
\caption{The sum of all distortion corrections. The sum
  includes the distortions caused  by the magnetic
  field  non-uniformities,  misalignment   between  the  axis  of  the
  magnetic  and  electric fields,  the  effects  of  a tilted  central
  membrane, non-flat end-caps,  and local electric field imperfections
  at the  junction of  the inner and  outer sectors at  $R\approx 120$
  cm.}
\label{distort}
\end{figure}

Figure \ref{distort} shows the sum  of the distortion corrections as a
function of radius  and $z$ inside the active volume  of the TPC. With
these  distortion corrections  applied, the  relative error  between a
point and  the track-model fit is  50 $\mu$m while  the absolute error
for any one point is about 500 $\mu$m.

\begin{table}[htb]
\begin{tabular}{|p{6.5cm}|p{3.1cm}|p{2.6cm}|}
\hline
Cause of the Distortion  & Magnitude of the Imperfection &
Magnitude of the Correction \\
\hline
Non-uniform B field &   $\pm 0.0040$ T &        0.10 cm \\
Geometrical effect between the inner and outer sub-sectors & 
Exact calculation based on geometry  & 0.05 cm (near pad row 13) \\
Cathode - non-flat shape and tilt   &   0.1 cm &        0.04 cm \\
The angular offset between E and B field &    0.2 mr & 0.03 cm \\
TPC endcaps - non-flat shape and tilt & 0.1 cm &        0.03 cm \\
Misalignment between IFC and OFC &      0.1 cm &        0.03 cm \\
Space Charge build up in the TPC &      0.001  C / $\epsilon_0$ & 0.03 cm average over volume \\
\hline
\end{tabular}
\vspace{0.2cm}
\caption{The distortion corrections applied to STAR data; their cause,
and the magnitude of their effect on the data.}
\label{tdistort}
\end{table}

\subsection{Two hit resolution}

\begin{figure}[htb]
\includegraphics[width=14cm]{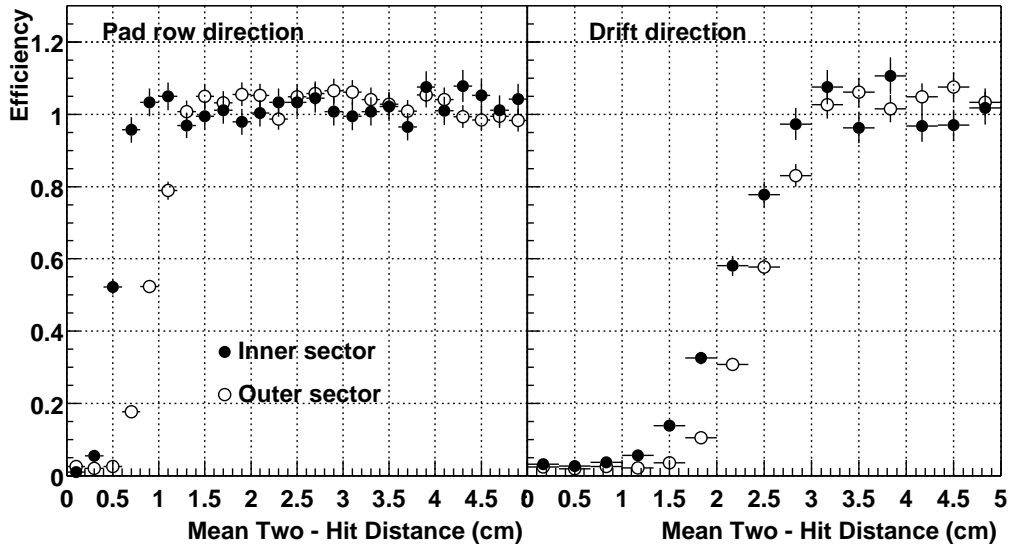}
\caption{Two-hit resolution in the STAR TPC.  The drift direction is
  along the $z$ axis and the  pad row direction is along the local $x$
  axis.}
\label{ftwohit}
\end{figure}

The inner and outer sub-sectors  have different size pads and so their
two-hit  resolutions are  different.  Figure  \ref{ftwohit}  shows the
efficiency  of  finding  two  hits  as  a  function  of  the  distance
separating them.  The efficiency  depends on whether the track segment
is observed in  the inner or the outer  sub-sectors. The efficiency is
the ratio of the distributions  of the distance separating 2 hits from
the same  event and  2 hits  from different events.   Two hits  can be
completely resolved  when they are  separated in the  padrow direction
(i.e. along the local $x$ axis) by at least 0.8 cm in the inner sector
and 1.3  cm in the outer  sector.  Similarly, two  hits are completely
resolved when  they are separated  in the drift direction  (i.e. along
the $z$ axis)  by 2.7 cm in the  inner sector and 3.2 cm  in the outer
sector.

\subsection{Tracking Efficiency}

The  tracking  software  performs  two  distinct  tasks.   First,  the
algorithms associate space points to form tracks and, second, they fit
the points on  a track with a track-model  to extract information such
as the momentum of the particle. The track-model is, to first order, a
helix.  Second order effects include  the energy lost in the gas which
causes a  particle trajectory to  deviate slightly from the  helix. In
this section,  we will discuss  the efficiency of finding  tracks with
the software.

The tracking efficiency depends on the acceptance of the detector, the
electronics detection  efficiency, as  well as the  two-hit separation
capability of the  system. The acceptance of the TPC  is 96\% for high
momentum  tracks  traveling   perpendicular  the  beamline.   The  4\%
inefficiency is  caused by  the spaces between  the sectors  which are
required to mount the wires on the sectors.  The software also ignores
any space  points that fall  on the  last 2 pads  of a pad  row.  This
fiducial  cut is  applied to  avoid position  errors that  result from
tracks not having  symmetric pad coverage on both  sides of the track.
It also  avoids possible local  distortions in the drift  field.  This
fiducial cut reduces the total acceptance to 94\%.

The  detection  efficiency of  the  electronics  is essentially  100\%
except for dead  channels and the dead channel  count is usually below
1\% of the total.  However,  the system cannot always separate one hit
from two  hits on adjacent pads  and this merging of  hits reduces the
tracking efficiency. The  software also applies cuts to  the data. For
example, a  track is  required to have  hits on  at least 10  pad rows
because shorter  tracks are too  likely to be broken  track fragments.
But this  cut can also remove  tracks traveling at a  small angle with
respect to the beamline and low momentum particles that curl up in the
magnetic field.   Since the merging  and minimum pad rows  effects are
non-linear, we can't do a simple calculation to estimate their effects
on the data.  We can simulate them, however.

In  order to  estimate  the tracking  efficiency,  we embed  simulated
tracks  inside real  events and  then  count the  number of  simulated
tracks that  are in the  data after the track  reconstruction software
has done  its job.   The technique allows  us to account  for detector
effects and especially the losses related to a high density of tracks.
The  simulated tracks  are very  similar to  the real  tracks  and the
simulator tries  to take into account  all the processes  that lead to
the detection of particles  including: ionization, electron drift, gas
gain, signal  collection, electronic amplification,  electronic noise,
and dead channels.  The results of the embedding studies indicate that
the systematic error on the tracking efficiency is about 6\%.

\begin{figure}[htb]
\includegraphics[width=14cm]{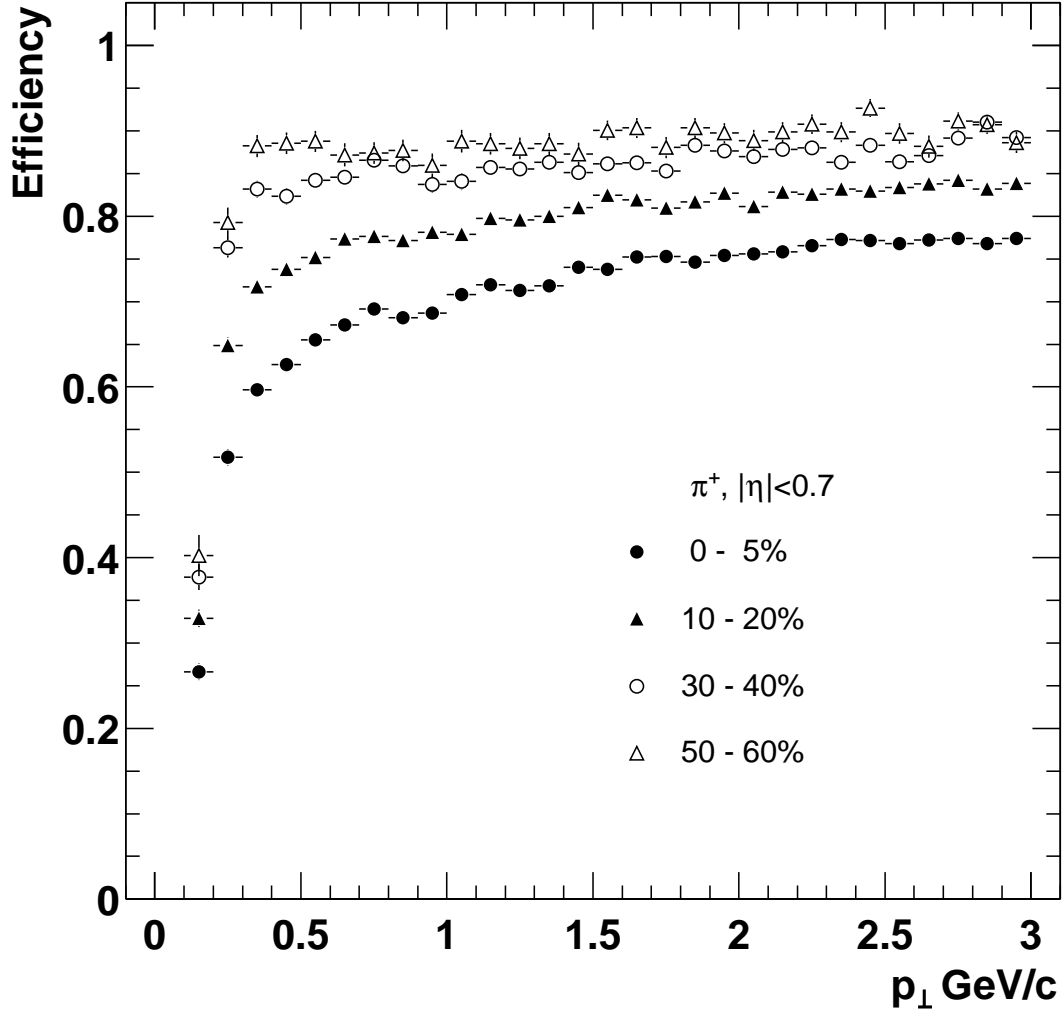}
\caption{The pion tracking efficiency in STAR for central Au+Au events
at RHIC.  Tracks  with $|y|<0.7$ were used to  generate the figure and
the  magnetic  field was  set  to  0.25 T.   The  data  are binned  by
centrality.   The   most  central  collisions  are   the  the  highest
multiplicity  data and  they  are  shown as  black  dots.  The  lowest
multiplicity data are shown as open triangles.}
\label{fefficiency}
\end{figure}

Figure \ref{fefficiency}  shows the pion  reconstruction efficiency in
Au+Au collisions  with different multiplicities  as a function  of the
transverse  momentum  of the  primary  particle\cite{manuel}. In  high
multiplicity  events it  reaches  a  plateau of  80\%  for high  $p_T$
particles.  Below  300 MeV/c the efficiency drops  rapidly because the
primary particles spiral  up inside the TPC and  don't reach the outer
field cage.  In addition, these  low momentum particles  interact with
the beam  pipe and the inner  field cage before  entering the tracking
volume of the TPC.  As a function of mulitplicity, the efficiency goes
up to the geometrical limit, minus software cuts, for low multiplicity
events.

\subsection{Vertex resolution}

The primary vertex can used  to improve the momentum resolution of the
tracks and  the secondary vertices  can be separated from  the primary
vertices if the vertex resolution  is good enough. Many of the strange
particles produced in heavy ion collisions can be identified this way.

\begin{figure}[htb]
\includegraphics[width=14cm]{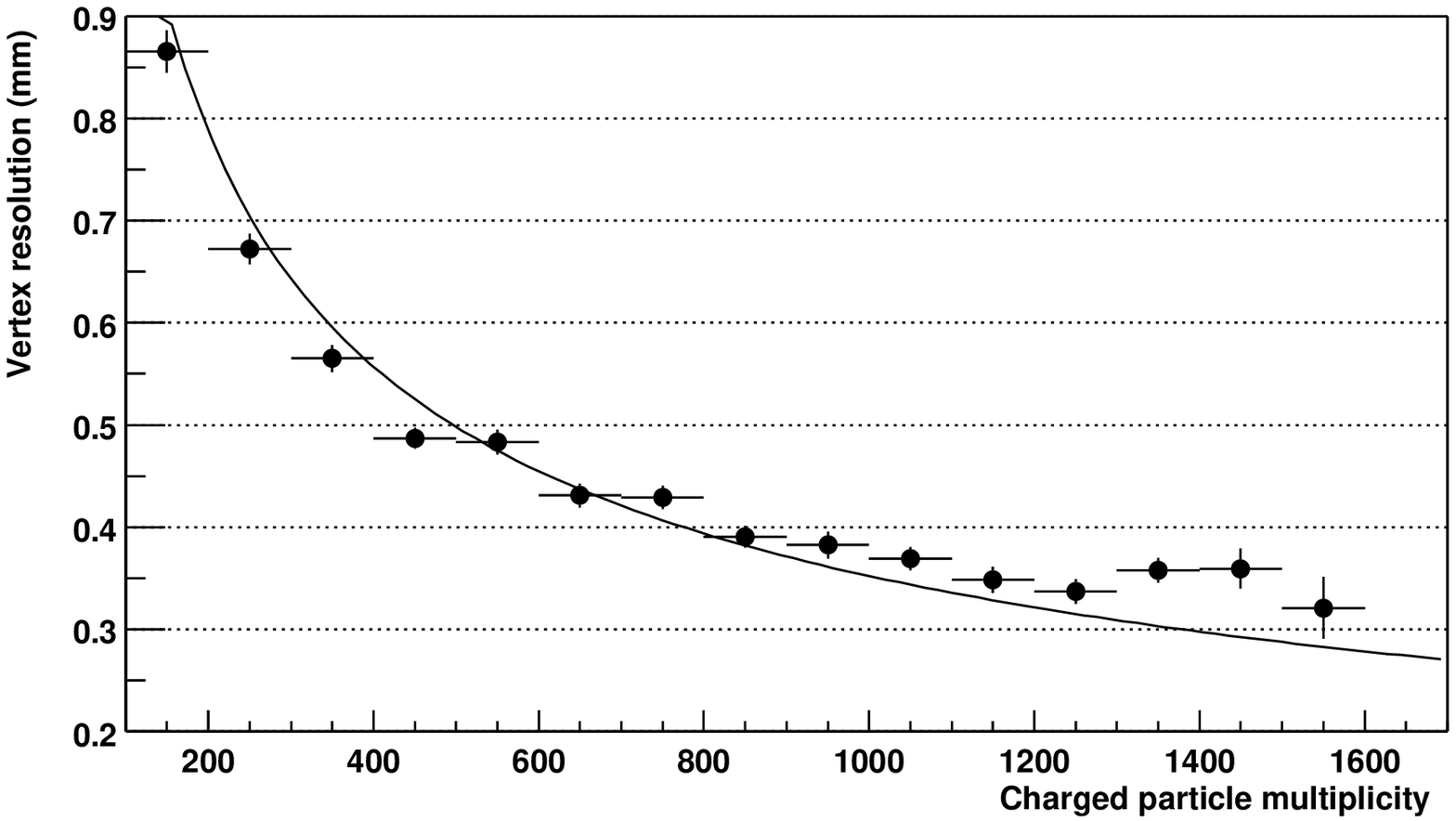}
\caption{Primary vertex resolution in the transverse plane.}
\label{fvertex}
\end{figure}

The  primary  vertex  is  found  by  considering  all  of  the  tracks
reconstructed  in the  TPC and  then  extrapolating them  back to  the
origin.  The global average is the vertex position. The primary vertex
resolution  is  shown in  Fig.  \ref{fvertex}.   It  is calculated  by
comparing the  position of the  vertices that are  reconstructed using
each  side  of  the  TPC,  separately.  As  expected,  the  resolution
decreases  as the  square root  of the  number of  tracks used  in the
calculation.  A resolution  of 350 $\mu$m is achieved  when there 
are more than 1,000 tracks.
 
\subsection{Momentum resolution}

The transverse momentum, $p_T$, of  a track is determined by fitting a
circle through the  $x$, $y$ coordinates of the  vertex and the points
along the track. The total momentum is calculated using this radius of
curvature and the  angle that the track makes with  respect to the $z$
axis of the TPC. This procedure works for all primary particles coming
from the vertex, but for secondary decays, such as $\Lambda$ or $K_s$,
the circle fit must be done without reference to the primary vertex.
 
\begin{figure}[htb]
\includegraphics[width=14cm]{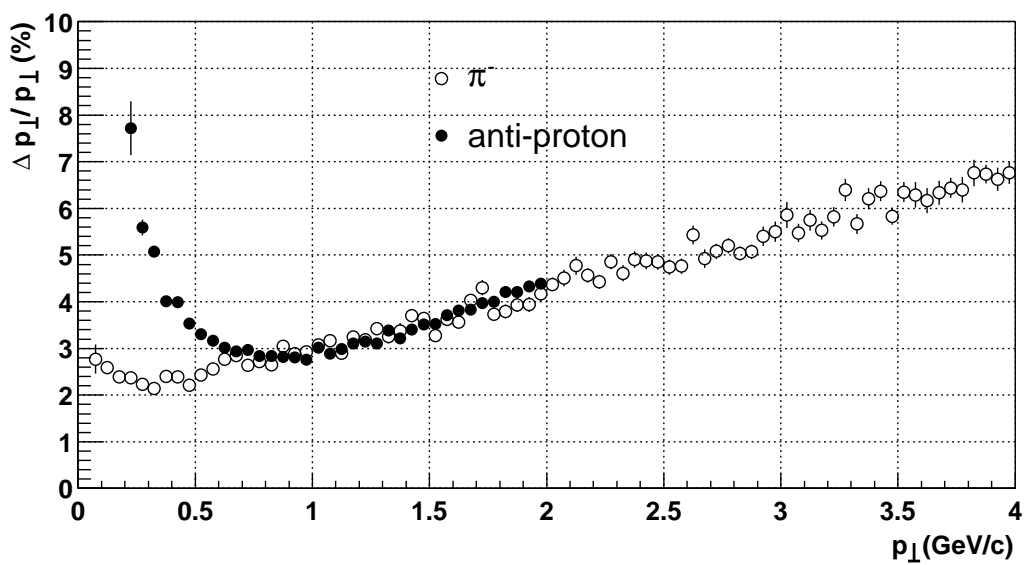}
\caption{Transverse momentum  resolution of  the STAR TPC  for $\pi^-$
and anti-protons in the 0.25  T magnetic field. Tracks are required to
be formed by  more than 15 hits.  Tracks are  embedded in minimum bias
events.  The  momentum  resolution   is  calculated  as  the  Gaussian
$sigma$.}
\label{fresolution}
\end{figure}

In  order to  estimate the  momentum resolution  we use  the embedding
technique discussed above.   The track simulator was used  to create a
track with  a known momentum.  The  track was then embedded  in a real
event in order to simulate the momentum smearing effects of working in
a high  track density environment. Figure  \ref{fresolution} shows the
$p_T$  resolution for  $\pi^-$ and  anti-protons in  STAR.  The figure
shows two regimes: at  low momentum, where multiple Coulomb scattering
dominates (i.e. $p_T < 400 MeV/c$ for pions, and $p_T < 800$ MeV/c for
anti-protons), and at higher momentum where the momentum resolution is
limited  by the  strength  of the  magnet  field and  the TPC  spatial
resolution. The best relative  momentum resolution falls between these
two extremes and it is 2\% for pions.

\subsection{Particle identification using $dE/dx$}

Energy lost in the TPC gas is a valuable tool for identifying particle
species. It  works especially well  for low momentum particles  but as
the particle energy rises, the energy loss becomes less mass-dependent
and it is  hard to separate particles with  velocities $v>0.7$c.  STAR
was designed to be able to separate pions and protons up to 1.2 GeV/c.
This  requires a relative $dE/dx$ resolution  of 7\%.   The challenge,
then,  is to  calibrate the  TPC and  understand the  signal  and gain
variations well enough to be able to achieve this goal.

The measured $dE/dx$  resolution depends on the gas  gain which itself
depends  on the  pressure in  the TPC.   Since the  TPC is  kept  at a
constant 2  mbar above atmospheric  pressure, the TPC  pressure varies
with time.  We monitor the gas  gain with a wire chamber that operates
in the  TPC gas return  line. It measures  the gain from  an $^{55}$Fe
source.  It will be used to  calibrate the 2001 data, but for the 2000
run, this  chamber was not installed  and so we monitored  the gain by
averaging the signal for tracks over the entire volume of the detector
and we  have done a relative  calibration on each sector  based on the
global  average. Local  gas  gain  variations   are  calibrated  by
calculating  the average signal  measured on  one row  of pads  on the
pad-plane and assuming that all pad-rows measure the same signal.  The
correction is done on the pad-row level because the anode wires lie on
top of, and run the full length of, the pad-rows.

The read-out  electronics also  introduce uncertainties in  the $dE/dx$
signals.  There are small variations between pads, and groups of pads,
due to the different response of each readout board.  These variations
are monitored by  pulsing the ground plane of the  anode and pad plane
read-out system and  then assuming that the response  will be the same
on every pad.

The $dE/dx$  is extracted from  the energy loss  measured on up  to 45
padrows.  The length  over which the particle energy  loss is measured
(pad  length modulo  the  crossing and  dip  angles) is  too short  to
average out  ionizations fluctuations.  Indeed,  particles lose energy
going through  the gas in frequent  collisions with atoms  where a few
tens of eV are released, as well as, rare collisions where hundreds of
eV  are  released~\cite{bichsel1}.   Thus,   it  is  not  possible  to
accurately measure  the average  $dE/dx$.  Instead, the  most probable
energy  loss  is  measured.   We  do  this  by  removing  the  largest
ionization  clusters.   The truncated  mean,  where  a given  fraction
(typically  30\%)  of  the  clusters  having the  largest  signal  are
removed, is  an efficient tool  to measure the most  probable $dE/dx$.
However,  fitting  the  $dE/dx$  distribution including  all  clusters
associated with a given track was found to be more effective.  It also
allows us  to account  for the variation  of the most  probable energy
loss with the length of the ionization samples ($dx$)~\cite{bichsel2}.

\begin{figure}[htb]
\includegraphics[width=14cm]{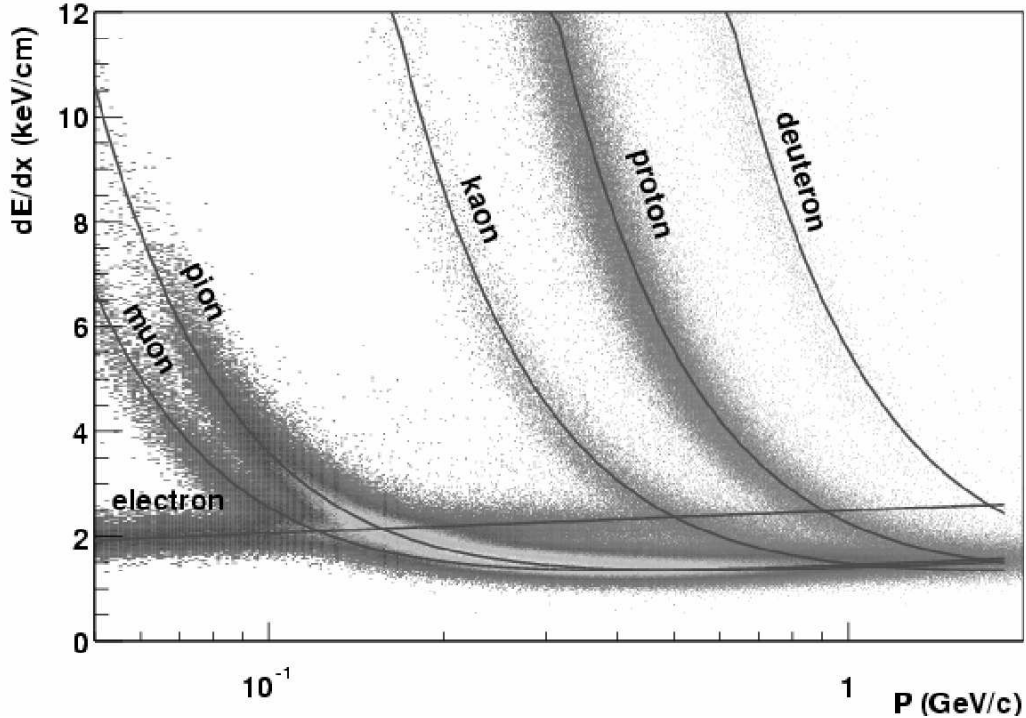}
\caption{The  energy  loss   distribution  for  primary  and  seconday
particles in  the STAR TPC as a  function of the $p_T$  of the primary
particle.  The magnetic field was 0.25 T.}
\label{fdedx}
\end{figure}

Figure \ref{fdedx} shows the energy loss for particles in the TPC as a
function of the  particle momentum.  The data have  been corrected for
signal  and gain  variations and  the data  are plotted  using  a 70\%
truncated mean. The  magnetic field setting is 0.25  T. The resolution
is 8\%  for a track  that crosses 40  pad-rows. At 0.5 T,  the $dE/dx$
resolution improves  because the  transverse diffusion is  smaller and
this  improves the  signal to  noise  ratio for  each cluster.  Figure
\ref{fdedx}  includes  both  primary  and  secondary  particles.   The
prominent  proton,  deuteron,  and  muon  bands  come  from  secondary
interactions in the beam pipe and  IFC, and from pion and kaon decays.
Pions and protons can be separated from each other up to 1 GeV/c.

\section{Conclusions}

The STAR  TPC is up  and running at  RHIC.  The detector  finished its
second year  of operation on January  25th, 2002 and  the operation of
the  TPC was  stable and  reliable  throughout both  run cycles.   Its
performance is very close to the original design requirements in terms
of   tracking  efficiency,  momentum   resolution,  and   energy  loss
measurements.  Many results from  the 2000/2001 data have already been
published and  they demonstrate that  the physics at RHIC  is exciting
and     rich.      We      invite     you     to     examine     these
papers\cite{star1,star2,star3,star4,star5,star6,star7}.

\section{Acknowledgements}

We  wish to thank  the RHIC  Operations Group  and the  RHIC Computing
Facility at  Brookhaven National  Laboratory, and the  National Energy
Research  Scientific Computing  Center at  Lawrence  Berkeley National
Laboratory for their support. This  work was supported by the Division
of  Nuclear Physics and  the Division  of High  Energy Physics  of the
Office of Science of the  U.S. Department of Energy, the United States
National  Science Foundation, the  Bundesministerium fuer  Bildung und
Forschung of  Germany, the Institut National de  la Physique Nucleaire
et  de  la Physique  des  Particules  of  France, the  United  Kingdom
Engineering and Physical Sciences Research Council, Fundacao de Amparo
a Pesquisa  do Estado  de Sao Paulo,  Brazil, the Russian  Ministry of
Science  and  Technology, the  Ministry  of  Education  of China,  the
National Natural Science Foundation of China, and the Swedish National
Science Foundation.

\end{document}